\documentclass[journal=jacsat,manuscript=article]{achemso}

\usepackage[version=3]{mhchem}

\newcommand{\beginsupplement}{%
        \setcounter{table}{0}
        \renewcommand{\thetable}{S\arabic{table}}%
        \setcounter{figure}{0}
        \renewcommand{\thefigure}{S\arabic{figure}}%
     }

\author{Ashutosh Patri}
\email{ashutosh.patri@polymtl.ca}
\affiliation[Polytechnique Montr\'eal]
{Department of Electrical Engineering, Polytechnique Montr\'eal, Montr\'eal}
\author{St\'{e}phane K\'{e}na-Cohen}
\email{s.kena-cohen@polymtl.ca}
\affiliation[Polytechnique Montr\'eal]
{Department of Engineering Physics, Polytechnique Montr\'eal, Montr\'eal, Montr\'eal}
\author{Christophe Caloz}
\email{christophe.caloz@polymtl.ca}
\affiliation[Polytechnique Montr\'eal]
{Department of Electrical Engineering, Polytechnique Montr\'eal, Montr\'eal}

\title {Large-Angle, Broadband and Multifunctional Gratings Based on Directively Radiating Waveguide Scatterers}


\begin{document}

\begin{abstract}
  
  Conventional surface-relief gratings are inefficient at deflecting normally-incident light by large angles. This constrains their use in many applications and limits the overall efficiency of any optical instrument integrating gratings. Here, we demonstrate a simple approach for the design of diffraction gratings that can be highly efficient for large deflection angles, while also offering additional functionality. The gratings are composed of a unit cell comprising a vertically-oriented asymmetric slot-waveguide. The unit cell shows oscillating unidirectional scattering behavior that can be precisely tuned as a function of the waveguide length. This occurs due to interference between multiple modes excited by the incident light. In contrast to metasurface-based gratings with multiple resonant sub-elements, a periodic arrangement of such non-resonant diffracting elements allows for broadband operation and a strong tolerance for variations in angle of incidence. Full-wave simulations show that our grating designs can exhibit diffraction efficiencies ranging from 94\% for a deflection angle of 47$^\circ$ to 80\% for deflection angle of 80$^\circ$.  To demonstrate the multifunctionality of our grating design technique, we have also proposed a flat polarization beamsplitter, which allows for the separation of the two orthogonal polarizations by 80$^\circ$, with an efficiency of 80\%.
\end{abstract}

\section{Introduction}
Optical systems rely on various components to manipulate the phase, amplitude, and polarization of light. For more than a century, diffraction gratings have been used to disperse polychromatic light or to deflect monochromatic light into desired directions. They find widespread use across diverse applications such as spectroscopy~\cite{roscoe1885spectrum}, ultrafast optics~\cite{treacy1969optical}, imaging~\cite{dobson1997diffractive}, optical holography~\cite{brown1969computer}, and integrated photonics~\cite{dakss1970grating}. To avoid signal loss into undesired diffraction orders, blazed gratings are commonly used. Blazed gratings, however, only operate efficiently for small deflection angles. For example, when the incoming light is normally-incident on a blazed transmission grating, the efficiency drops below 80\% at angles larger than only~$\sim$20$^\circ$~\cite{swanson1991binary, lalanne1999waveguiding}. This restricts their use in several areas such as flat lens imaging~\cite{fujita1982blazed} and high-resolution spectroscopy~\cite{loewen1997diffraction}. The inefficiency of blazed gratings comes from the shadowing effect~\cite{swanson1991binary, lalanne1998blazed} inherent to their sawtooth topology. One solution is the use of binary-blazed gratings where each grating period consists of multiple sub-elements acting as discrete phase-shifters. The waveguiding nature of each sub-element helps eliminate the shadowing effect by concentrating the field inside the dielectric structure~\cite{lalanne1999waveguiding}. However, for deflection angles larger than~$\sim$40$^\circ$, it becomes difficult to adequately discretize the phase elements within a grating period and simultaneously avoid coupling between adjacent waveguides~\cite{lalanne1999design, lalanne2017metalenses}. The super-modes in the coupled-waveguide assembly cause difficulties in phase-control and gives rise to energy propagation outside the waveguide structure, which in turn leads to energy leakage into undesired diffraction orders. Moreover, high aspect-ratio structures are often required for adequate phase discretization that cannot be fabricated precisely and these inaccuracies further reduce diffraction efficiency.

In contrast, by relying on resonant phase delay rather than propagational phase delay, as in surface-relief gratings, metasurface-based gratings do not exhibit shadowing effect nor do they require high aspect-ratio structures. Metasurface-gratings use subwavelength-sized plasmonic~\cite{meinzer2014plasmonic} or dielectric~\cite{kuznetsov2016optically} resonators and therefore are far more flexible than natural materials in terms of phase and polarization control. The deep-subwavelength size of plasmonic resonators help in fine discretization of the phase  within the diffraction period, but their efficiency is limited due to ohmic metal losses at optical frequencies~\cite{khurgin2012reflecting}. On the other hand, dielectric resonators have negligible absorptive loss, but due to the unavailability of optical-materials with very high dielectric constant, their relatively large lateral dimensions severely restrict the adequate phase discretization required for large deflection angles. To avoid the complexity involved in the phase discretization process, Ref.~\citenum{sell2017large} introduced a robust optimization technique to realize large-angle diffraction gratings using a single dielectric resonator within each diffraction period. Subsequently, in Ref.~\citenum{khaidarov2017asymmetric} the physical mechanism behind such resonators was described, along with an intuitive design example, in terms of a directional nanoantenna with multipole interactions. A more quantitative description for multipole interactions in bianosotropic resonators based grating was recently proposed in Ref.~\citenum{fan2018perfect}. However, due to the resonant behaviour of these grating elements, metasurface-gratings suffer from narrow operational bandwidth and unstable operation for small variations in the angle of incidence~\cite{lalanne2017metalenses}.

In this work, we propose a different and simpler strategy for realizing efficient gratings for large deflection angles,  which we refer to as Directive Waveguide Scatterer Gratings (DWSG)~\cite{dwsg}. These use a single \emph{non-resonating} directive scattering element within the diffraction period. The low aspect ratio waveguide scatterer, integrated in our design, can replace the multiple sub-elements required for binary-blazed gratings or metasurface-gratings. As an all-dielectric grating, DWSGs have negligible absorption losses as compared to plasmonic metasurface-based gratings. In addition, unlike dispersive dielectric resonator based gratings, the non-resonant nature of the DWSG element ensures a large operational bandwidth and weak sensitivity to variations in the angle of incidence. We describe the physical mechanism behind the waveguide-based approach, which relies on the control of the scattering pattern using interference between multiple modes excited by the incident beam. We demonstrate gratings with efficiencies $>80$\% for deflection angles ranging from 40$^\circ$--80$^\circ$ using one and two-dimensional arrays of asymmetric slot-waveguides. Finally, to demonstrate the polarization control capability of our gratings, we also show a design for a polarization beamsplitter based on the DWSG concept.

\section{DWSG Designs and Results}

\begin{figure}[b]
\centering 
\includegraphics[width=1\textwidth]{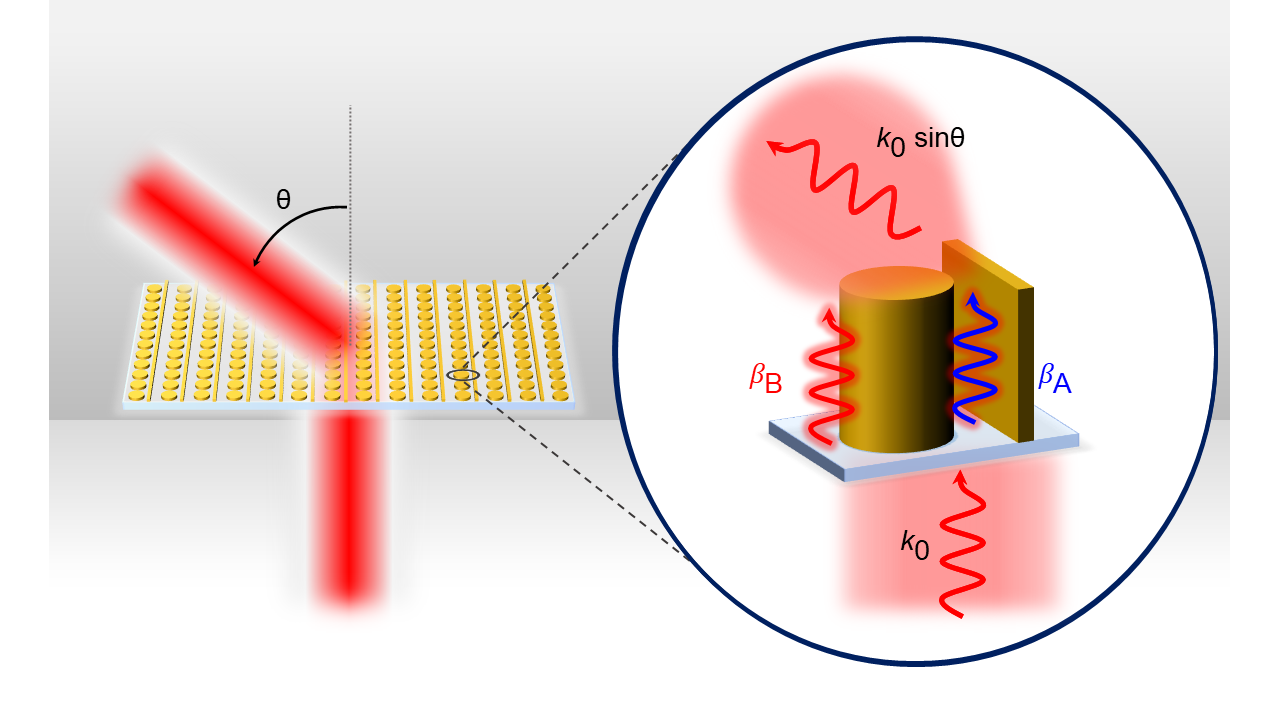}
\caption{A schematic illustration of the Directive Waveguide Scatterer Grating (DWSG). The grating deflects a normally-incident beam to a desired direction $\theta$, while suppressing all undesired diffraction orders. The inset depicts a unit cell comprising an asymmetric slot-waveguide structure. When illuminated by an incident light with propagation constant $k_\text{0}$, it radiates the maximum amount of power along the $k_\text{0}\text{sin}\theta$ direction. This is due to the interference between a guided mode with propagation constant $\beta_\text{A}$, and a higher-order guided or radiation mode with most of its energy propagating in the free-space region, with propagation constant $\beta_\text{B}$. Both modes are excited by the incident light.} \label{fig:illustration}
\end{figure}

\begin{figure}[b]
\centering 
\includegraphics[width=1\textwidth]{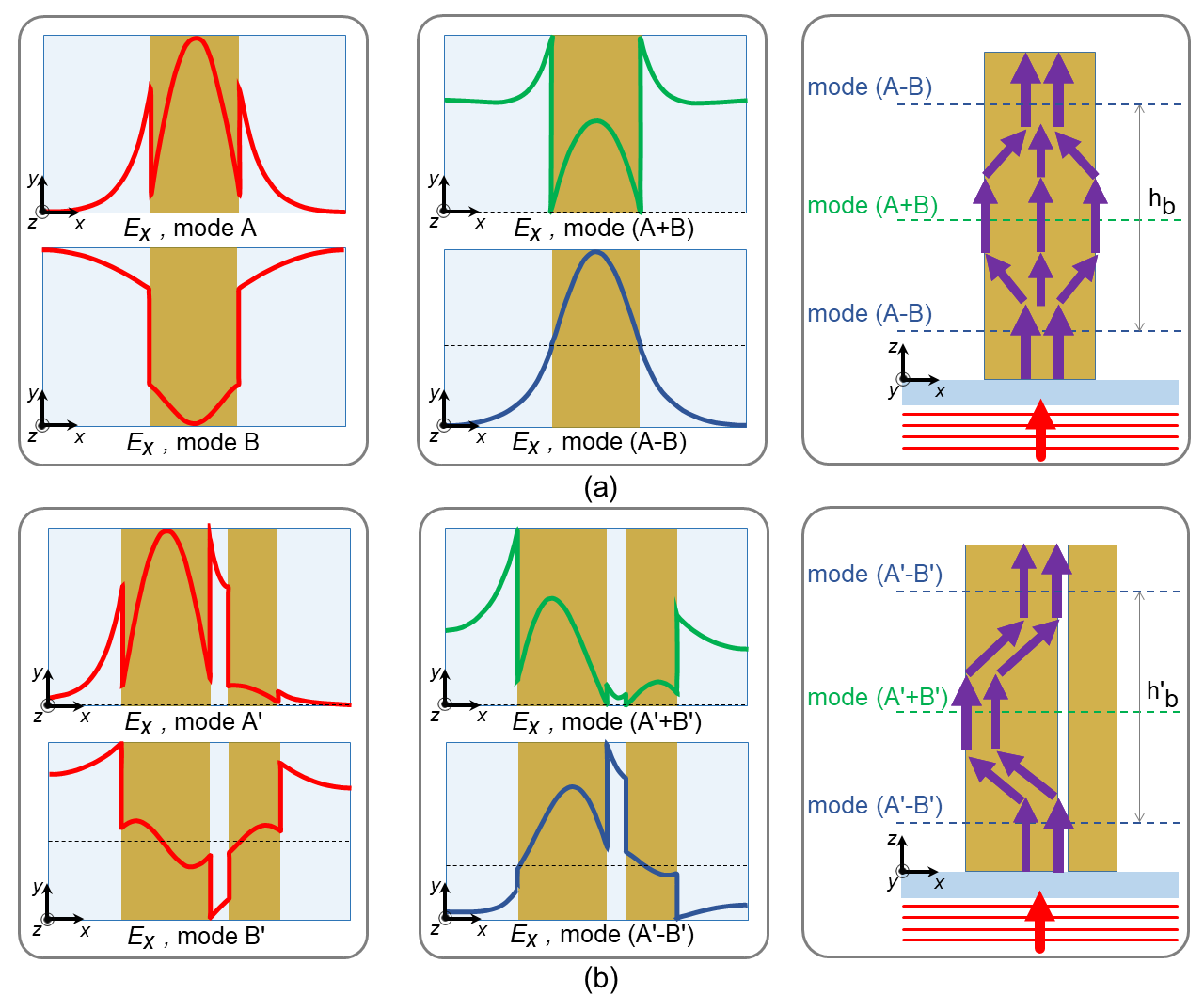}
\caption{An illustration of the interference effect governing the radiation pattern of the DWSG unit cell. (a) A unit cell comprising a symmetric waveguide. (b) A unit cell comprising an asymmetric slot-waveguide. The left panels shows the transverse electric field profile for the two dominant modes excited by the incident plane wave i.e. \emph{mode A} and \emph{mode B} in the case of the symmetric unit cell and \emph{mode A'} and \emph{mode B'} in the case of the asymmetric unit cell. The black dashed line corresponds to $E_{x}=0$. \emph{Mode A} and \emph{mode A'} are the fundamental modes of the corresponding waveguides, whereas \emph{mode B} and \emph{mode B'} are a radiation mode and higher-order guided mode supported by the unit cell, respectively, with effective indices close to $n_\text{air}$. The difference in propagation constants of \emph{mode A} and \emph{mode B} or \emph{mode A'} and \emph{mode B'} leads to interference. The center panels show the superposition of the two modes for constructive and destructive interference. The position of these interference patterns along the height of the unit cell is shown in the right panel with dashed green (constructive) and blue (destructive) lines. The red arrow in the right panel represents the Poynting vector for the incident plane wave and the purple lines represent the Poynting vectors along the height of the structure for the combined field generated due to the interference. The beating period is denoted as $h_\text{b}$ and $h'_\text{b}$ for symmetric and asymmetric unit cell, respectively.} \label{fig:interference}
\end{figure}

The key objective underlying DWSG design is to redirect the maximum amount of incident power into a desired diffraction order while suppressing all undesired ones. For example, Figure~\ref{fig:illustration} illustrates a schematic of DWSG illuminated from the substrate side. The DWSG diffracts the entire incident light energy in the direction of the desired diffraction order, chosen here to be $m=-1$ in transmission. For an intuitive understanding, we can consider that the diffraction grating consists of an array of secondary radiators excited by a primary excitation source corresponding to the incident beam. The overall scattering pattern from the array is then a function of the element factor, consisting of the radiation pattern of each individual secondary element, and the array factor, representing the radiation pattern of an array of isotropically radiating elements. The advantage of this analysis method, governed by pattern multiplication principle~\cite{schelkunoff1943mathematical}, is that it allows to treat the element and array factor separately during the design process. In contrast to the equivalent Bloch mode analysis~\cite{sell2017large}, the pattern multiplication principle provides a straightforward understanding of the role of the array and its elements~\cite{khaidarov2017asymmetric}.

For an infinite number of isotropic scattering-elements, the array factor simply reduces to a Dirac delta function peaked in the same direction as the diffraction orders of a conventional grating~\cite{Haupt2010antenna}. For a given incident and deflection angle, the corresponding grating period $\Lambda_\text{d}$ can be calculated using the grating equation. For simplicity, we will consider a wavelength-scale period $\Lambda_\text{d}$ such that only the zero and first-order diffractions are supported ($m=-1,0,1$). This results in three propagating beams on the transmission side and three on the reflection side. To further simplify the analysis, we can then make two broad assumptions. First, the incident beam width is considered to be much larger than the array period, which allows us to approximate the light incident on individual scattering elements as a plane wave. Second, there is no change in orientation among the scattering elements, which allows us to consider a single polarization state. In this simplified scenario, the amount of power scattered into each diffraction order can then be controlled by changing the element factor, i.e. the scattering pattern of the secondary radiator or the unit cell. 

For the design of an efficient transmission grating, the objective is to align the maximum scattering direction of the secondary radiator with the direction of desired diffraction order on the transmission side, while aligning the null radiation directions with that of the undesired diffraction orders. For this purpose, we use a unit cell composed of a vertically-oriented waveguide section as the secondary radiator element. The plane wave illuminating the unit cell aperture can excite both guided and radiation modes in the structure. Because there is a single sub-wavelength thick waveguide within a wavelength-scale period, a considerable amount of energy will propagate in the free-space region adjacent to the waveguide. This contributes significantly to the excitation of higher-order guided modes and radiation modes. In practice, as illustrated in Figure~\ref{fig:interference}, a plane wave incident on the unit cell will mostly excite the lowest order guided mode of the waveguide (with an effective index close to that of the dielectric material) and a high order mode, which can be either guided or radiative, with an effective index approaching $n_\text{air}$~\cite{snyder2012optical, laakmann1976waveguides}. For symmetric waveguides, the excited modes will be those with even parity to match that of the incident plane wave. Modes in asymmetric waveguides do not have a well-defined parity, but in practice, those closer to even parity will be primarily excited. The total field at any given cross section along $z$-axis is then given by the vector sum of all the modes excited by the incident plane wave. Importantly, the difference in propagation constants for the modes with large excitation amplitudes will lead to beating and a consequent periodic variation of the lateral field distribution along $z$-axis. This gives rise to a zig-zag evolution of the energy flow along the height of the waveguide as illustrated in the right panels of Figure~\ref{fig:interference}. For the asymmetric waveguide, we can create an asymmetric scattering pattern with maximum scattering in the same direction as the desired diffraction order by choosing a specific waveguide height. Furthermore, if there is sufficient overlap between the incident field and modes excited in the unit cell, the diffraction orders in reflection can be completely suppressed. 

\begin{figure}[b]
\centering 
\includegraphics[width=1\textwidth]{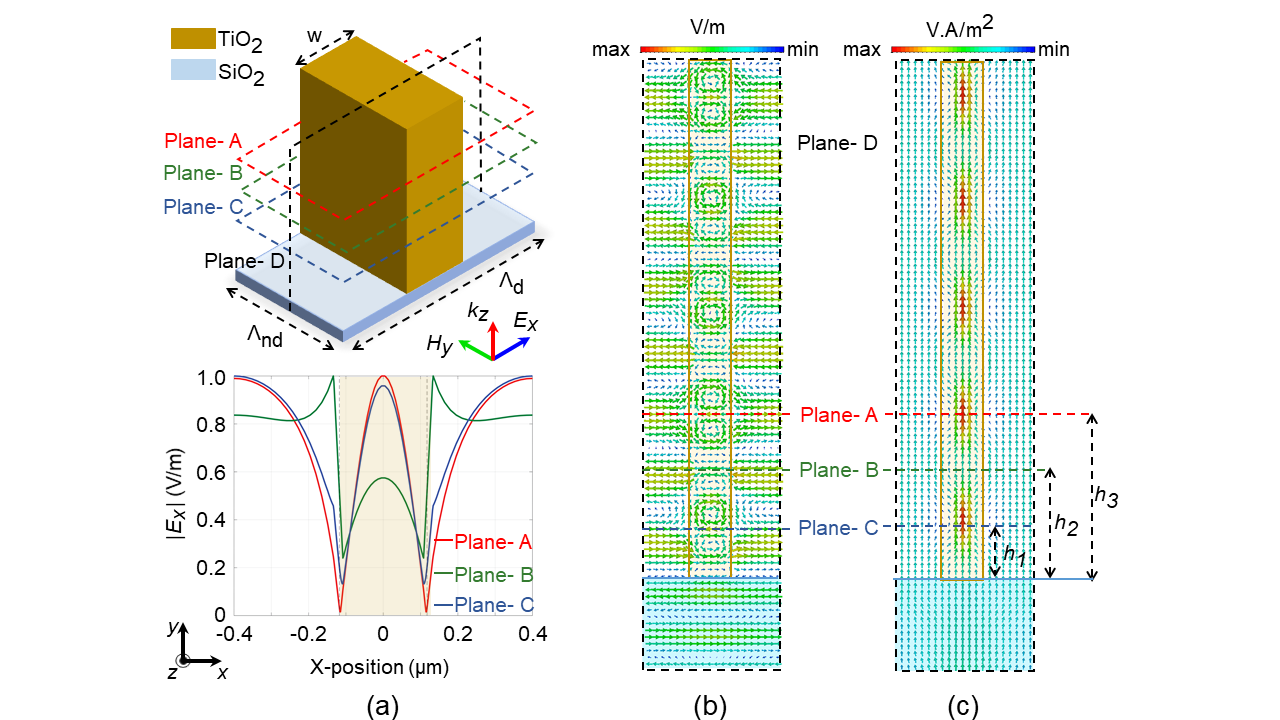}
\caption{Illustration of the field evolution within a one-dimensional array of symmetric waveguides. The design parameter values are, in nm : $\Lambda_\text{d}=800, \Lambda_\text{nd}=350,~ \text{and}~ w=235$. The half-beat length value is found to be $\sim$290 nm, for $\lambda=680$ nm, both from our 2D mode solver ($n_\text{eff, A}=2.14, n_\text{eff, B}=0.96$) and 3D full-wave simulation results. The horizontal cutting-plane A, B, and C are defined at a height of $h_{1}=290~\text{nm}, h_{2}=580~\text{nm},~\text{and}~h_{3}=870~\text{nm}, \text{respectively}$. (a) Perspective view of the unit cell (top) and $| E_{x}|$ component (bottom) on the intersection lines between Plane-A, B, and C with Plane-D. (b) Absolute $E$ field on Plane-D. (c) Absolute Poynting vector on Plane-D.} \label{fig:DWSG_symmetric_phy}
\end{figure}

We begin by quantitatively illustrating these ideas using a simple symmetric Ti$\text{O}_{2}$ waveguide as the scattering element. As shown in Figure~\ref{fig:DWSG_symmetric_phy}(a), we consider a one-dimensional arrangement of the waveguides that is diffractive only along one lateral dimension, i.e. the $x$-axis , with period $\Lambda_\text{d}$. The waveguide is continuous along the other lateral dimension, i.e. the $y$-axis. For the 3D simulation, we use a non-diffractive period $\Lambda_\text{nd}$ along this direction.
 The dimensions of the waveguide element are chosen such that the unit cell supports an even guided mode $\text{TM}_{0}$, an odd guided mode $\text{TM}_{1}$, and a radiation mode(see Figure~\ref{fig:modes_symmetric}). We first examine the variation in the lateral field distribution as a function of waveguide height. The incident plane wave predominantly excites the lowest order TM mode, mode A, with $n_\text{eff, A}$=2.14 and the even radiation mode, mode B, with $n_\text{eff, B}$=0.96 approaching $n_\text{air}$. To show the beating, we define three horizontal cutting-planes A, B, and C intersecting a vertical cross-section D of the unit cell at heights $h_{1}$, $h_{2}$, and $h_{3}$, respectively. For these three heights,  $| E_{x}|$ is shown in Figure~\ref{fig:DWSG_symmetric_phy}(a). The field intensity $| E_{x}|$ shows destructive interference between mode A and B at height $h_{2}$ and constructive interference at heights $h_{1}$ and $h_{3}$. This contrasts sharply with the unchanging lateral field profile that would be obtained upon exciting only the lowest order guided mode. The corresponding beat length $h_\text{b}=h_\text{3}-h_\text{1}= 2(h_\text{2}-h_\text{1})$ is simply given by
\begin{equation}\label{eqn:beat_length}
    h_\text{b}=\lambda_\text{0}/(n_\text{eff, A}-n_\text{eff, B}),
\end{equation}
where $\lambda_\text{0}$ is the free-space wavelength. It should be noted that the $n_\text{eff}$ value used in eq~\ref{eqn:beat_length} is for an infinitely extended waveguide section. In practice, the required height for our optimized grating designs will vary slightly from eq~\ref{eqn:beat_length} due to abruptly terminated open-ends.

The rate at which the fields vary due to beating depends on the difference between $n_\text{eff, A}$ and $n_\text{air}$ or more precisely $n_\text{eff, B}$. Given that the time-averaged power integrated over planes A, B and C is almost constant~\cite{radiationmode}, the Poynting vector inside the waveguide must alternate between pointing towards or away from the center as a function of the position along the $z$-axis to maintain the varying field profile. This is shown in Figure~\ref{fig:DWSG_symmetric_phy}(c) for the one-dimensional waveguide. Smaller beat lengths lead to a higher rate of field variation and create a steep energy gradient along the lateral direction, which allows for the Poynting vector components to have a larger angle with respect to the $z$-axis. Further discussion on the Poynting vector behavior will be given in the following sections. 

\begin{figure}[b]
\centering 
\includegraphics[width=1\textwidth]{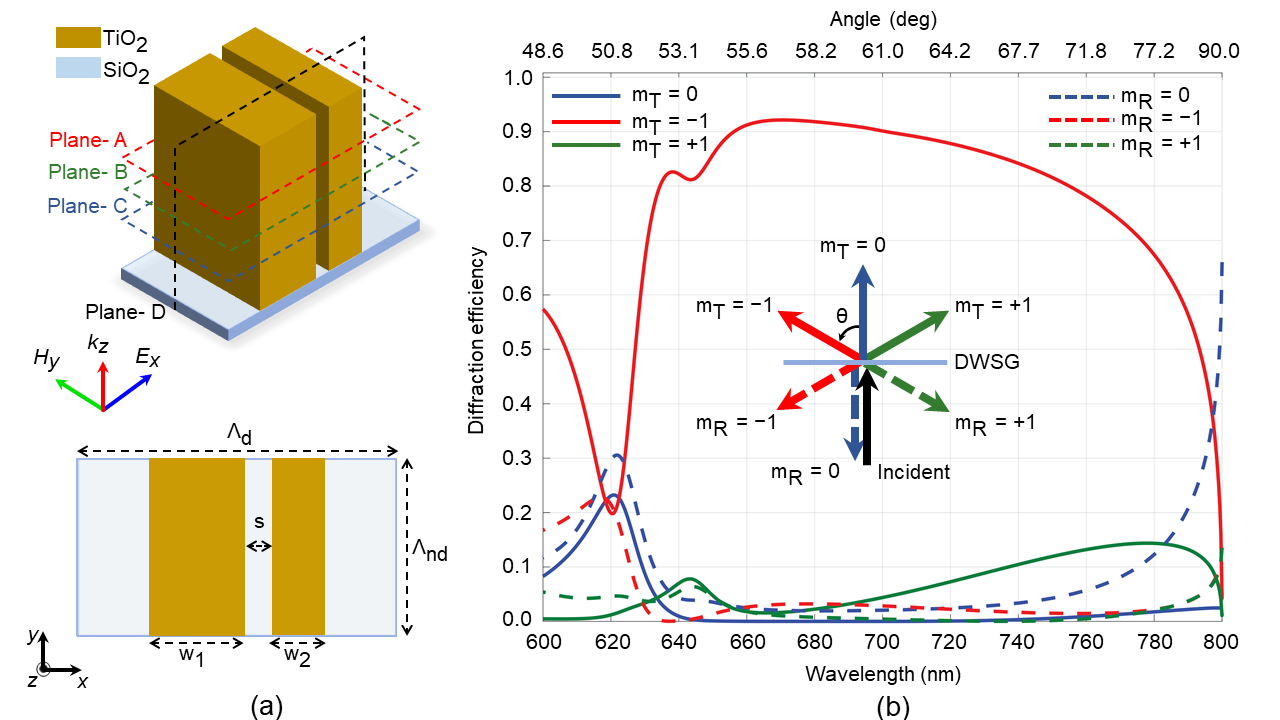}
\caption{Design of a unit cell comprising a one-dimensional asymmetric slot-waveguide and its grating efficiency. (a) Perspective view (top) and top view (bottom). (b) Diffraction efficiency as a function of wavelength for a 390 nm structural height. The design parameter values are, in nm : $\Lambda_\text{d}=800, \Lambda_\text{nd}=350, w_\text{1}=235, w_\text{2}=135, ~\text{and}~s=50$.} \label{fig:DWSG_asymmetric_des}
\end{figure}

Directing all of the incident power into a \emph{single} diffraction order requires an asymmetric scattering element. For design flexibility, we use an asymmetric slot-waveguide. We will demonstrate examples of both one-dimensional and two-dimensional arrays of asymmetric slot-waveguides, while highlighting their respective strengths. Figure~\ref{fig:DWSG_asymmetric_des}(a), shows a unit cell comprising a one-dimensional asymmetric slot-waveguide. For this example, we have simply added a thin dielectric slab next to the symmetric structure previously shown in Figure~\ref{fig:DWSG_symmetric_phy}(a). Figure~\ref{fig:DWSG_asymmetric_des}(b) shows the diffraction efficiency of the grating. This design leads to a high diffraction efficiency of 50\% for a large deflection angle of $\sim82.5^{\circ}$ and a maximum efficiency of 92\% for a deflection angle of $\sim57^{\circ}$. The simulation result shows a diffraction efficiency of more than 80\%, for deflection angles ranging from $\sim$52$^\circ$ to 70$^\circ$, over a $\sim$120 nm bandwidth covering the entire red spectrum.

\begin{figure}[b]
\centering 
\includegraphics[width=1\textwidth]{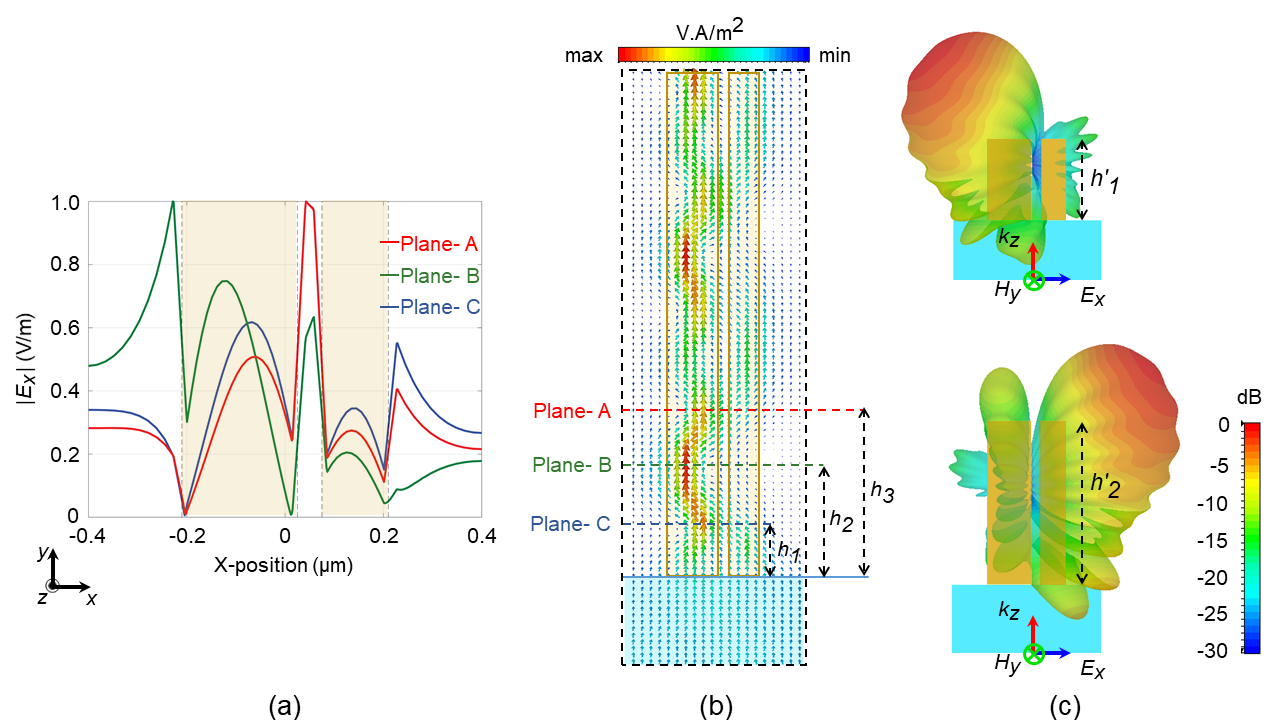}
\caption{Physical mechanism behind the scattering element presented in Figure~\ref{fig:DWSG_asymmetric_des}. The half beat length value is found to be $\sim$350 nm, for $\lambda=670$~nm, both from our 2D mode solver ($n_\text{eff, A}=2.02, n_\text{eff, B}=1.12$) and 3D full-wave simulation results. The horizontal cutting-plane A, B, and C are defined at a height of $h_{1}=280~\text{nm}, h_{2}=630~\text{nm},~\text{and}~h_{3}=980~\text{nm}, \text{respectively}$ to show the variations of the field. (a) $| E_{x}|$ component on the intersection lines between Plane-A, B, and C with Plane-D. (b) Absolute Poynting vector plot on Plane-D. (c) Directional radiation pattern for an open-ended waveguide section with $h'_{1}=440$ nm (top), and $h'_{2}=880$ nm (bottom).} 
\label{fig:DWSG_asymmetric_phy}
\end{figure}

The underlying mechanism behind the asymmetric scattering is nearly identical to that shown above, except for the asymmetric energy flow. The incident plane wave excites two dominant even guided modes~\cite{oddevenmode} ($\text{TM}_{0}~\text{and}~\text{TM}_{2}$, see Figure~\ref{fig:modes_slot}) and their interference leads to strong field variation along the height of the unit cell. The $| E_x|$ profiles along the horizontal cutting planes from Figure~\ref{fig:DWSG_asymmetric_des}(b) are shown in Figure~\ref{fig:DWSG_asymmetric_phy}(a). As before, we observe a strong variation of the $| E_x|$ component along the propagation direction. The effective refractive index for the asymmetric slot-waveguide, which dictates the beat length, can be readily obtained from the equations in Refs.~\citenum{ma2009analysis,  almeida2004guiding}. The asymmetric flow of the Poynting vectors is shown in Figure~\ref{fig:DWSG_asymmetric_phy}(b). In  Figure~\ref{fig:DWSG_asymmetric_phy}(c), we show that by chosing height $h'_\text{1}$ or $h'_\text{2}$, corresponding to asymmetric outward or inward flowing Poynting vectors, respectively, the asymmetric radiation pattern can be directed along either the $m=+1$ or $m=-1$ diffraction orders, while suppressing the undesired diffraction orders. We have also analyzed its response as a function of incident angle and the simulation results are shown in Figure~\ref{fig:anglescan}. A high diffraction efficiency for a broad range of incidence angles, along $-\theta$ (defined in glass), is observed. For $+\theta$ larger than $15^{\circ}$, the $m=-1$ diffraction order becomes evanescent.

To increase the diffraction efficiency for very large angles, a higher value of $n_\text{eff, A}$ is required to increase the lateral energy gradient. However, increasing the width of the slot-waveguide or replacing the waveguide material with a higher dielectric constant material, such as silicon, can cause significant reflection loss due to the large mismatch between the substrate and waveguide refractive indices. This can be remedied by using a multilayer architecture, with a Ti$\text{O}_{2}$ spacer to minimize the reflection from a silicon slot-waveguide (see Figure~\ref{fig:DWSG_asymmetric_bilayer}). A remarkably high diffraction efficiency of 50\% is observed for a deflection angle of 86$^\circ$ and an efficiency of more than 80\%, ranging from $\sim$70$^\circ$ to 80$^\circ$, over a bandwidth of 50 nm is obtained.

\begin{figure}[b]
\centering 
\includegraphics[width=1\textwidth]{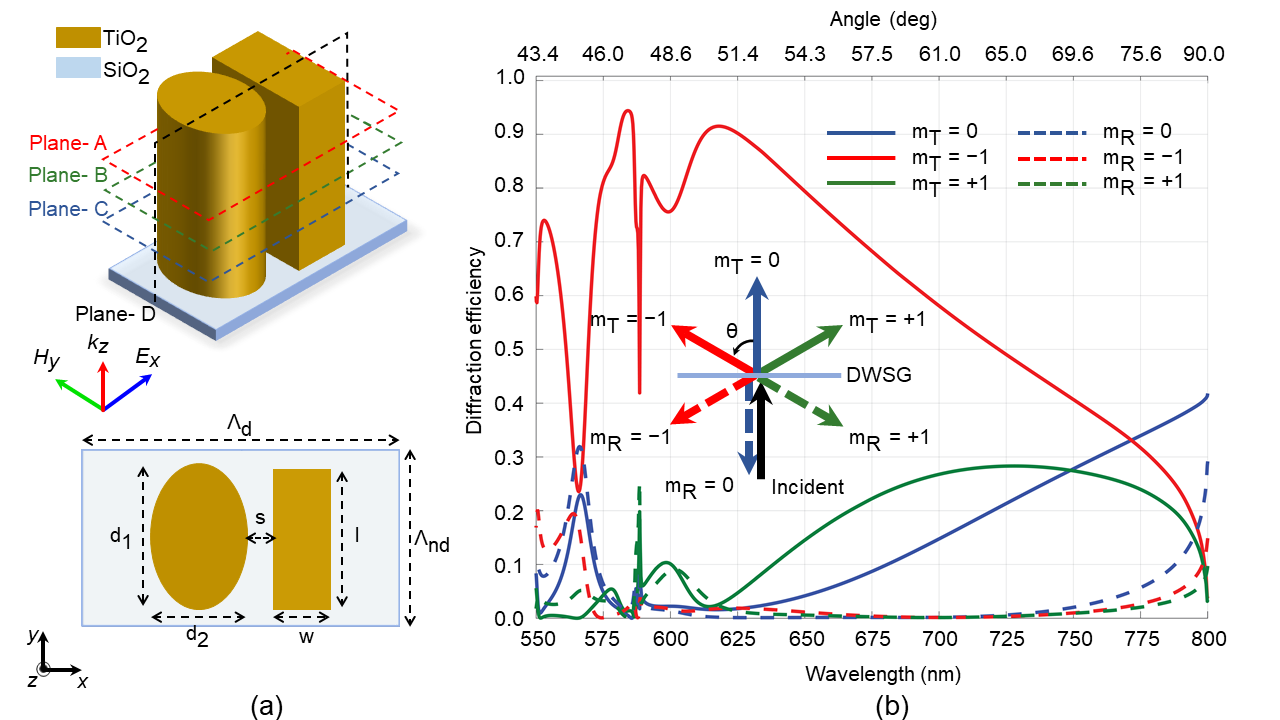}
\caption{Design of a unit cell comprising a two-dimensional asymmetric slot-waveguide and its grating efficiency. (a) Perspective view (top) and top view (bottom). (b) Diffraction efficiency as a function of wavelength for a 400 nm structural height. The design parameter values are, in nm : $\Lambda_\text{d}=800,~\Lambda_\text{nd}=350,~d_\text{1}=270,~d_\text{2}=210,~l=280,~w=125,~\text{and}~ s=50.$} \label{fig:DWSG_asymmetric_des_2D}
\end{figure}

\begin{figure}[b]
\centering 
\includegraphics[width=1\textwidth]{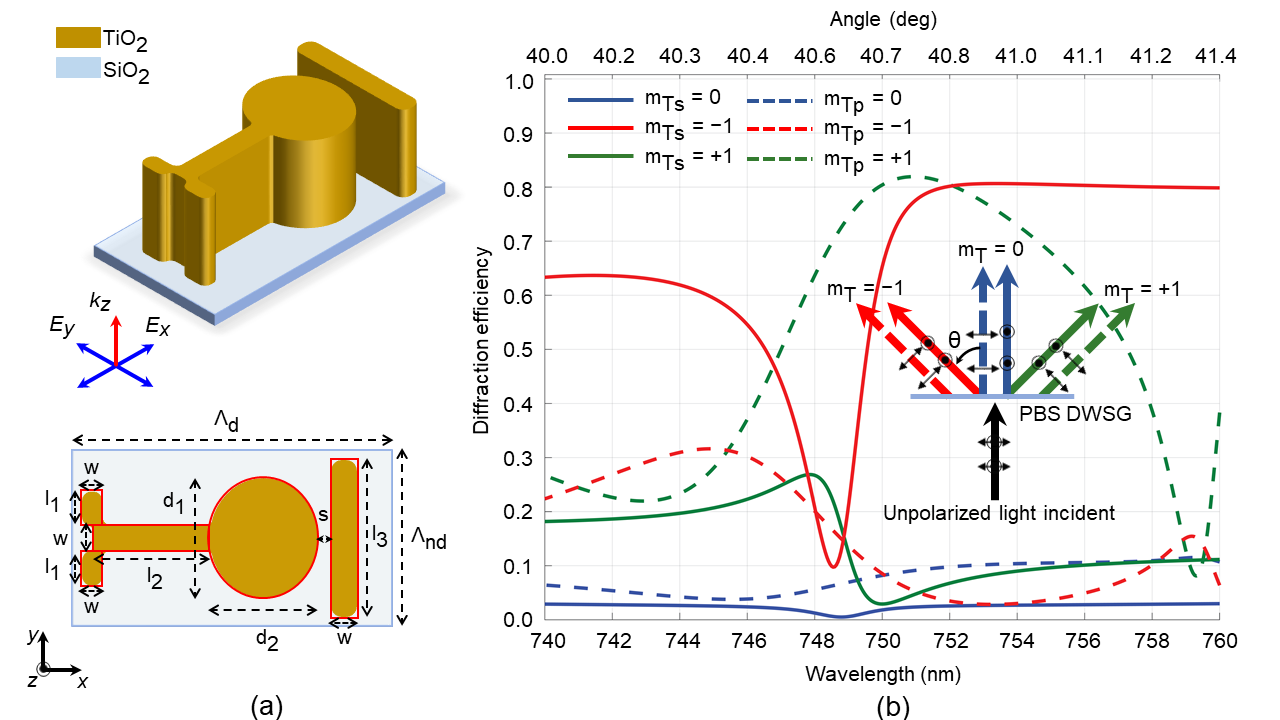}
\caption{Design of a unit cell comprising a two-dimensional asymmetric slot-waveguide and its polarization beamsplitter grating efficiency. (a)~Perspective view (top) and top view (bottom) of the unit cell with design parameters. (b)~Diffraction efficiency for transmitted diffraction orders as a function of wavelength. The parameter values are, in nm : $\Lambda_\text{d}=1150, \Lambda_\text{nd}=500, w=70, l_\text{1}=100,  l_\text{2}=430, l_\text{3}=450, d_\text{1}=340, d_\text{2}=380, ~\text{and}~ s=50$. All the parts of the structure have the same height of 390~nm. The inward or outward bends in the design are of 30~nm radius considering a practical electron-beam-lithography process. } \label{fig:DWSG_PBS_des}
\end{figure}

The two-dimensional slot-waveguide provides two advantages over its one-dimensional counterpart. First, it allows for an extra set of design parameters that can be adjusted for further optimization. Second, it allows for polarization control. In Figure~\ref{fig:DWSG_asymmetric_des_2D}(a), we demonstrate a two-dimensional asymmetric waveguide design for moderate deflection angles, ranging from $\sim45^{\circ}$ to $55^{\circ}$. The diffraction efficiency is shown in Figure~\ref{fig:DWSG_asymmetric_des_2D}(b) and it reaches a maximum of 94.4\% for a deflection angle of $\sim47^{\circ}$ and 91.64\% for $\sim50^{\circ}$. The eigenmodes of the unit cell aperture are shown in Figure~\ref{fig:modes_ellipse} and the variation of $|E_{x}|$ on various lateral planes, the oscillating flow of the Poynting vectors and the asymmetric directional radiation patterns are shown in Figure~\ref{fig:DWSG_asymmetric_phy_2D}. To demonstrate the polarization control capability of this DWSG, we designed a Polarization Beamsplitter (PBS) grating as shown in Figure~\ref{fig:DWSG_PBS_des}(a). The design was optimized from the asymmetric two-dimensional unit cell discussed in the previous example and it operates at a center wavelength of $\lambda_\text{0}=752$~nm. As shown in Figure~\ref{fig:DWSG_PBS_des}(b), it reaches an efficiency of $\sim80\%$ for both polarizations with a separation angle of $80^\circ$ between the polarized light beams. The radiation pattern of the unit cell, when illuminated with a $x$- and $y$-polarized plane waves from the substrate side, can be found in Figure~\ref{fig:DWSG_PBS_phy}. The polarization extinction ratio is $\sim12$ dB for the beams along both positive and negative first-order diffraction directions at $752$ nm.

\section{Discussion}

A variety of recent approaches for directional scattering have reported theoretical diffraction efficiencies of 75\%, 85\% and 95\% for moderate deflection angles of 50$^\circ$~\cite{sell2017large}, 53$^\circ$~\cite{khaidarov2017asymmetric} and 45$^\circ$~\cite{fan2018perfect}, respectively for resonant dielectric structure based ``metagratings". In contrast, non-resonant DWSGs exhibit efficiencies of 91.64\%, 92\% and 94.4\% for deflection angles of 50$^\circ$, 57$^\circ$ and 47$^\circ$, respectively. For very large angles such as 75$^\circ$ and 82$^\circ$, competing approaches have obtained theoretical efficiencies of 86\%~\cite{sell2017large} and 50\%~\cite{khaidarov2017asymmetric}, respectively. Meanwhile, the design presented here using the single layer one-dimensional DWSG from Figure~\ref{fig:DWSG_asymmetric_des} exhibits theoretical efficiencies of 73\% and 50\% and the multilayer DWSG (see Figure~\ref{fig:DWSG_asymmetric_bilayer}) shows efficiencies of 87\% and 73\%, respectively, for those deflection angles. For devices with multiple constraints, such as polarization-control in the PBS-DWSG, further robust optimization~\cite{sell2017large} could be used to improve the efficiency.

An inherent advantage of the non-resonant scattering is the larger bandwidth. For example, we have a 23\% fractional bandwidth, $\Delta\lambda/\lambda$, where $\Delta\lambda$ is the 3 dB bandwidth and $\lambda$ is the center wavelength, for the one-dimensional DWSG (Figure~\ref{fig:DWSG_asymmetric_des}), as compared to a $\sim$10\% bandwidth for the gratings reported in Ref.~\citenum{khaidarov2017asymmetric, fan2018perfect}. In comparison to our asymmetric one-dimensional DWSG, the asymmetric two-dimensional DWSG (Figure~\ref{fig:DWSG_asymmetric_des_2D}) provides the advantages of polarization manipulation, but with the expense of a reduced fractional bandwidth of 20\% due to the presence of the sharp resonance peak, at $\lambda=588$~nm, corresponding to the excitation of in-plane guided mode. The non-diffractive period in the two-dimensional DWSG does not support coupling to any free-space modes, however it can support coupling to in-plane guided modes due the smaller effective wavelength. 

Our incident angle sensitivity analysis shows that the efficiency is robust with respect to changes of the incident angle along the $-\theta$ direction. Given that most of the energy is tightly confined near the sub-wavelength slot region, the response does not vary significantly with changes of the in-plane wavevector, $k_{||}$. This is similar to the case of binary blazed gratings~\cite{lalanne2017metalenses}, where most of the energy in concentrated in the dielectric region of the waveguide. In contrast, most dielectric resonator based metasurface-gratings~\cite{lalanne2017metalenses} show strong spatial dispersion and are thus not suitable for variable incident angles.

In the case of multiple non-resonant sub-element based gratings~\cite{lalanne1999waveguiding,pietarinen2008double, arbabi2015dielectric, chen2018broadband}, the phase-shifting elements require a height of $\lambda_{0}/(n_\text{max}-1)$, where the $n_{max}$ is the maximum effective refractive index of the thickest waveguide element. In contrast, DWSG structures require an approximate height of $\lambda_\text{0}/2(n_{\text{eff}}-1)$ where $n_{\text{eff}}$ is the lowest order effective index of the asymmetric waveguide. The $n_{\text{eff}}$ in latter will generally be larger than $n_\text{max}$ due to the use of wider waveguides, which allows for the use of low aspect-ratio structures. The complex structures reported in the literature for large deflection angles of 80$^\circ$~\cite{khaidarov2017asymmetric} can be prone to fabrication errors, however, and the corresponding experimental efficiency was found to be less than 10\%. In contrast, our one-dimensional DWSG architectures are relatively simple and we have limited the slot width to a minimum of 50 nm in all our examples to minimize fabrication complexity.

In conclusion, we have presented a new paradigm for the design of high-performance diffraction gratings, which are both multifunctional and flexible with respect to the diffraction angles. We have demonstrated, using simulation, a variety of designs based on the use of asymmetric slot- waveguide scatterers that radiate in a controllable directional fashion due to the interference between multiple guided modes and radiation modes. DWSGs can be implemented using simple structures and show encouraging results for large-angle deflections,  providing  absolute diffraction efficiency of more  than  80\%  for deflection angles ranging from $\sim$40$^\circ$ to 80$^\circ$. We have shown that our non-resonant scatterer based design strategy can be used to make gratings which are insensitive to variations in incident angle and exhibit large operational bandwidth required for various practical applications. Moreover, this principle can be further applied for design of reflective gratings and partially reflective-transmissive gratings. We believe DWSGs can bridge the gap between  conventional  diffraction  gratings  and  metasurface-gratings,  by  combining the best of both i.e. simplicity and ease of fabrication of conventional gratings and improved performance and versatility of metasurface-gratings.

\section{Methods}
 We calculated the $n_\text{eff}$ values of the various modes excited in the unit cell structure from 2D frequency domain solver in COMSOL Multiphysics and the resulting beat length values were verified from 3D time domain solver in CST Studio. The unit cell design, both in 2D solver in COMSOL and 3D solver in CST, have two parallel Perfect Electric Conductor (PEC) and two parallel Perfect Magnetic Conductor (PMC) boundary conditions for the four lateral boundaries and Perfectly Matched Layer (PML) boundary conditions for the top and bottom boundaries in the 3D design. The beat length calculated from the previous steps provided the structural height required for our initial design. Then, we optimized the design parameters of the waveguide to achieve suitable diffraction efficiency in CST Studio Frequency domain solver using Periodic Boundary Conditions (PBC). 
 
 Absolute diffraction efficiency was computed for light incident from the glass substrate side (refractive index $n = 1.50$) and the transmitted beams, after diffraction, are in air ($n=1$). The waveguides are composed of Ti$\text{O}_{2}$ with $n=2.53$ for mono-layer DWSG and with an additional layer of Si with $n=3.5$ for bi-layer DWSG.

\begin{acknowledgement}

A.P. thanks Oscar V. Cespedes, Guillaume Lavigne, Muhammad Mohsin, Louis Haeberle, and Xiao Jia for valuable discussions. S.K.C. and C.C. acknowledge support from the Canada Research Chairs program and funding for this work from NSERC Strategic Grant STPGP-506808.

\end{acknowledgement}

\bibliography{achemso-demo}

\providecommand{\latin}[1]{#1}
\providecommand*\mcitethebibliography{\thebibliography}
\csname @ifundefined\endcsname{endmcitethebibliography}
  {\let\endmcitethebibliography\endthebibliography}{}
\begin{mcitethebibliography}{31}
\providecommand*\natexlab[1]{#1}
\providecommand*\mciteSetBstSublistMode[1]{}
\providecommand*\mciteSetBstMaxWidthForm[2]{}
\providecommand*\mciteBstWouldAddEndPuncttrue
  {\def\EndOfBibitem{\unskip.}}
\providecommand*\mciteBstWouldAddEndPunctfalse
  {\let\EndOfBibitem\relax}
\providecommand*\mciteSetBstMidEndSepPunct[3]{}
\providecommand*\mciteSetBstSublistLabelBeginEnd[3]{}
\providecommand*\EndOfBibitem{}
\mciteSetBstSublistMode{f}
\mciteSetBstMaxWidthForm{subitem}{(\alph{mcitesubitemcount})}
\mciteSetBstSublistLabelBeginEnd
  {\mcitemaxwidthsubitemform\space}
  {\relax}
  {\relax}

\bibitem[Roscoe and Schuster(1885)Roscoe, and Schuster]{roscoe1885spectrum}
Roscoe,~H.~E.; Schuster,~A. \emph{Spectrum analysis}; Macmillan, 1885\relax
\mciteBstWouldAddEndPuncttrue
\mciteSetBstMidEndSepPunct{\mcitedefaultmidpunct}
{\mcitedefaultendpunct}{\mcitedefaultseppunct}\relax
\EndOfBibitem
\bibitem[Treacy(1969)]{treacy1969optical}
Treacy,~E. \emph{IEEE Journal of quantum Electronics} \textbf{1969}, \emph{5},
  454--458\relax
\mciteBstWouldAddEndPuncttrue
\mciteSetBstMidEndSepPunct{\mcitedefaultmidpunct}
{\mcitedefaultendpunct}{\mcitedefaultseppunct}\relax
\EndOfBibitem
\bibitem[Dobson \latin{et~al.}(1997)Dobson, Sun, and
  Fainman]{dobson1997diffractive}
Dobson,~S.~L.; Sun,~P.-c.; Fainman,~Y. \emph{Applied optics} \textbf{1997},
  \emph{36}, 4744--4748\relax
\mciteBstWouldAddEndPuncttrue
\mciteSetBstMidEndSepPunct{\mcitedefaultmidpunct}
{\mcitedefaultendpunct}{\mcitedefaultseppunct}\relax
\EndOfBibitem
\bibitem[Brown and Lohmann(1969)Brown, and Lohmann]{brown1969computer}
Brown,~B.; Lohmann,~A. \emph{IBM Journal of research and Development}
  \textbf{1969}, \emph{13}, 160--168\relax
\mciteBstWouldAddEndPuncttrue
\mciteSetBstMidEndSepPunct{\mcitedefaultmidpunct}
{\mcitedefaultendpunct}{\mcitedefaultseppunct}\relax
\EndOfBibitem
\bibitem[Dakss \latin{et~al.}(1970)Dakss, Kuhn, Heidrich, and
  Scott]{dakss1970grating}
Dakss,~M.; Kuhn,~L.; Heidrich,~P.; Scott,~B. \emph{Applied physics letters}
  \textbf{1970}, \emph{16}, 523--525\relax
\mciteBstWouldAddEndPuncttrue
\mciteSetBstMidEndSepPunct{\mcitedefaultmidpunct}
{\mcitedefaultendpunct}{\mcitedefaultseppunct}\relax
\EndOfBibitem
\bibitem[Swanson(1991)]{swanson1991binary}
Swanson,~G.~J. \emph{Binary optics technology: theoretical limits on the
  diffraction efficiency of multilevel diffractive optical elements};
  1991\relax
\mciteBstWouldAddEndPuncttrue
\mciteSetBstMidEndSepPunct{\mcitedefaultmidpunct}
{\mcitedefaultendpunct}{\mcitedefaultseppunct}\relax
\EndOfBibitem
\bibitem[Lalanne(1999)]{lalanne1999waveguiding}
Lalanne,~P. \emph{JOSA A} \textbf{1999}, \emph{16}, 2517--2520\relax
\mciteBstWouldAddEndPuncttrue
\mciteSetBstMidEndSepPunct{\mcitedefaultmidpunct}
{\mcitedefaultendpunct}{\mcitedefaultseppunct}\relax
\EndOfBibitem
\bibitem[Fujita \latin{et~al.}(1982)Fujita, Nishihara, and
  Koyama]{fujita1982blazed}
Fujita,~T.; Nishihara,~H.; Koyama,~J. \emph{Optics letters} \textbf{1982},
  \emph{7}, 578--580\relax
\mciteBstWouldAddEndPuncttrue
\mciteSetBstMidEndSepPunct{\mcitedefaultmidpunct}
{\mcitedefaultendpunct}{\mcitedefaultseppunct}\relax
\EndOfBibitem
\bibitem[Loewen and Popov(1997)Loewen, and Popov]{loewen1997diffraction}
Loewen,~E.~G.; Popov,~E. \emph{Diffraction gratings and applications}; CRC
  Press, 1997\relax
\mciteBstWouldAddEndPuncttrue
\mciteSetBstMidEndSepPunct{\mcitedefaultmidpunct}
{\mcitedefaultendpunct}{\mcitedefaultseppunct}\relax
\EndOfBibitem
\bibitem[Lalanne \latin{et~al.}(1998)Lalanne, Astilean, Chavel, Cambril, and
  Launois]{lalanne1998blazed}
Lalanne,~P.; Astilean,~S.; Chavel,~P.; Cambril,~E.; Launois,~H. \emph{Optics
  letters} \textbf{1998}, \emph{23}, 1081--1083\relax
\mciteBstWouldAddEndPuncttrue
\mciteSetBstMidEndSepPunct{\mcitedefaultmidpunct}
{\mcitedefaultendpunct}{\mcitedefaultseppunct}\relax
\EndOfBibitem
\bibitem[Lalanne \latin{et~al.}(1999)Lalanne, Astilean, Chavel, Cambril, and
  Launois]{lalanne1999design}
Lalanne,~P.; Astilean,~S.; Chavel,~P.; Cambril,~E.; Launois,~H. \emph{JOSA A}
  \textbf{1999}, \emph{16}, 1143--1156\relax
\mciteBstWouldAddEndPuncttrue
\mciteSetBstMidEndSepPunct{\mcitedefaultmidpunct}
{\mcitedefaultendpunct}{\mcitedefaultseppunct}\relax
\EndOfBibitem
\bibitem[Lalanne and Chavel(2017)Lalanne, and Chavel]{lalanne2017metalenses}
Lalanne,~P.; Chavel,~P. \emph{Laser \& Photonics Reviews} \textbf{2017},
  \emph{11}, 1600295\relax
\mciteBstWouldAddEndPuncttrue
\mciteSetBstMidEndSepPunct{\mcitedefaultmidpunct}
{\mcitedefaultendpunct}{\mcitedefaultseppunct}\relax
\EndOfBibitem
\bibitem[Meinzer \latin{et~al.}(2014)Meinzer, Barnes, and
  Hooper]{meinzer2014plasmonic}
Meinzer,~N.; Barnes,~W.~L.; Hooper,~I.~R. \emph{Nature Photonics}
  \textbf{2014}, \emph{8}, 889--898\relax
\mciteBstWouldAddEndPuncttrue
\mciteSetBstMidEndSepPunct{\mcitedefaultmidpunct}
{\mcitedefaultendpunct}{\mcitedefaultseppunct}\relax
\EndOfBibitem
\bibitem[Kuznetsov \latin{et~al.}(2016)Kuznetsov, Miroshnichenko, Brongersma,
  Kivshar, and Luk’yanchuk]{kuznetsov2016optically}
Kuznetsov,~A.~I.; Miroshnichenko,~A.~E.; Brongersma,~M.~L.; Kivshar,~Y.~S.;
  Luk’yanchuk,~B. \emph{Science} \textbf{2016}, \emph{354}, aag2472\relax
\mciteBstWouldAddEndPuncttrue
\mciteSetBstMidEndSepPunct{\mcitedefaultmidpunct}
{\mcitedefaultendpunct}{\mcitedefaultseppunct}\relax
\EndOfBibitem
\bibitem[Khurgin and Boltasseva(2012)Khurgin, and
  Boltasseva]{khurgin2012reflecting}
Khurgin,~J.~B.; Boltasseva,~A. \emph{MRS bulletin} \textbf{2012}, \emph{37},
  768--779\relax
\mciteBstWouldAddEndPuncttrue
\mciteSetBstMidEndSepPunct{\mcitedefaultmidpunct}
{\mcitedefaultendpunct}{\mcitedefaultseppunct}\relax
\EndOfBibitem
\bibitem[Sell \latin{et~al.}(2017)Sell, Yang, Doshay, Yang, and
  Fan]{sell2017large}
Sell,~D.; Yang,~J.; Doshay,~S.; Yang,~R.; Fan,~J.~A. \emph{Nano letters}
  \textbf{2017}, \emph{17}, 3752--3757\relax
\mciteBstWouldAddEndPuncttrue
\mciteSetBstMidEndSepPunct{\mcitedefaultmidpunct}
{\mcitedefaultendpunct}{\mcitedefaultseppunct}\relax
\EndOfBibitem
\bibitem[Khaidarov \latin{et~al.}(2017)Khaidarov, Hao, Paniagua-Dom{\'\i}nguez,
  Yu, Fu, Valuckas, Yap, Toh, Ng, and Kuznetsov]{khaidarov2017asymmetric}
Khaidarov,~E.; Hao,~H.; Paniagua-Dom{\'\i}nguez,~R.; Yu,~Y.~F.; Fu,~Y.~H.;
  Valuckas,~V.; Yap,~S. L.~K.; Toh,~Y.~T.; Ng,~J. S.~K.; Kuznetsov,~A.~I.
  \emph{Nano letters} \textbf{2017}, \emph{17}, 6267--6272\relax
\mciteBstWouldAddEndPuncttrue
\mciteSetBstMidEndSepPunct{\mcitedefaultmidpunct}
{\mcitedefaultendpunct}{\mcitedefaultseppunct}\relax
\EndOfBibitem
\bibitem[Fan \latin{et~al.}(2018)Fan, Shcherbakov, Allen, Allen, Wenner, and
  Shvets]{fan2018perfect}
Fan,~Z.; Shcherbakov,~M.~R.; Allen,~M.; Allen,~J.; Wenner,~B.; Shvets,~G.
  \emph{ACS Photonics} \textbf{2018}, \emph{5}, 4303--4311\relax
\mciteBstWouldAddEndPuncttrue
\mciteSetBstMidEndSepPunct{\mcitedefaultmidpunct}
{\mcitedefaultendpunct}{\mcitedefaultseppunct}\relax
\EndOfBibitem
\bibitem[dws()]{dwsg}
Since the term ``meta'' generally refers to homogenizable structures, the
  terminology ``metagrating'' used by others is not appropriate here. In
  contrast, DWSG specifically and unambiguously describes the physics of the
  structure that suppresses all the diffraction orders except the one that is
  desired.\relax
\mciteBstWouldAddEndPunctfalse
\mciteSetBstMidEndSepPunct{\mcitedefaultmidpunct}
{}{\mcitedefaultseppunct}\relax
\EndOfBibitem
\bibitem[Schelkunoff(1943)]{schelkunoff1943mathematical}
Schelkunoff,~S.~A. \emph{The Bell System Technical Journal} \textbf{1943},
  \emph{22}, 80--107\relax
\mciteBstWouldAddEndPuncttrue
\mciteSetBstMidEndSepPunct{\mcitedefaultmidpunct}
{\mcitedefaultendpunct}{\mcitedefaultseppunct}\relax
\EndOfBibitem
\bibitem[Haupt(2010)]{Haupt2010antenna}
Haupt,~R.~L. \emph{Antenna arrays: a computational approach}; John Wiley \&
  Sons, 2010\relax
\mciteBstWouldAddEndPuncttrue
\mciteSetBstMidEndSepPunct{\mcitedefaultmidpunct}
{\mcitedefaultendpunct}{\mcitedefaultseppunct}\relax
\EndOfBibitem
\bibitem[Snyder and Love(2012)Snyder, and Love]{snyder2012optical}
Snyder,~A.~W.; Love,~J. \emph{Optical waveguide theory}; Springer Science \&
  Business Media, 2012\relax
\mciteBstWouldAddEndPuncttrue
\mciteSetBstMidEndSepPunct{\mcitedefaultmidpunct}
{\mcitedefaultendpunct}{\mcitedefaultseppunct}\relax
\EndOfBibitem
\bibitem[Laakmann and Steier(1976)Laakmann, and Steier]{laakmann1976waveguides}
Laakmann,~K.~D.; Steier,~W.~H. \emph{Applied optics} \textbf{1976}, \emph{15},
  1334--1340\relax
\mciteBstWouldAddEndPuncttrue
\mciteSetBstMidEndSepPunct{\mcitedefaultmidpunct}
{\mcitedefaultendpunct}{\mcitedefaultseppunct}\relax
\EndOfBibitem
\bibitem[rad()]{radiationmode}
The radiation mode is weakly guiding and thus decays in the spatially transient
  region where we observe the beating effect. In the spatially steady-state
  region, only the guided modes exist. Because the height of the waveguides
  considered in our examples are a few wavelengths ($\lambda_{0}$) long, the
  decay can be considered negligible and the power in each lateral plane
  constant.\relax
\mciteBstWouldAddEndPunctfalse
\mciteSetBstMidEndSepPunct{\mcitedefaultmidpunct}
{}{\mcitedefaultseppunct}\relax
\EndOfBibitem
\bibitem[odd()]{oddevenmode}
Although modes in an asymmetric waveguides are not strictly even or odd, we can
  define an even mode when the transverse field components have same polarity
  outside the both extreme edges of the waveguide and odd modes when they have
  opposite polarity.\relax
\mciteBstWouldAddEndPunctfalse
\mciteSetBstMidEndSepPunct{\mcitedefaultmidpunct}
{}{\mcitedefaultseppunct}\relax
\EndOfBibitem
\bibitem[Ma \latin{et~al.}(2009)Ma, Zhang, and Van~Keuren]{ma2009analysis}
Ma,~C.; Zhang,~Q.; Van~Keuren,~E. \emph{Optics Communications} \textbf{2009},
  \emph{282}, 324--328\relax
\mciteBstWouldAddEndPuncttrue
\mciteSetBstMidEndSepPunct{\mcitedefaultmidpunct}
{\mcitedefaultendpunct}{\mcitedefaultseppunct}\relax
\EndOfBibitem
\bibitem[Almeida \latin{et~al.}(2004)Almeida, Xu, Barrios, and
  Lipson]{almeida2004guiding}
Almeida,~V.~R.; Xu,~Q.; Barrios,~C.~A.; Lipson,~M. \emph{Optics letters}
  \textbf{2004}, \emph{29}, 1209--1211\relax
\mciteBstWouldAddEndPuncttrue
\mciteSetBstMidEndSepPunct{\mcitedefaultmidpunct}
{\mcitedefaultendpunct}{\mcitedefaultseppunct}\relax
\EndOfBibitem
\bibitem[Pietarinen and Vallius(2008)Pietarinen, and
  Vallius]{pietarinen2008double}
Pietarinen,~J.; Vallius,~T. \emph{Optics express} \textbf{2008}, \emph{16},
  13824--13830\relax
\mciteBstWouldAddEndPuncttrue
\mciteSetBstMidEndSepPunct{\mcitedefaultmidpunct}
{\mcitedefaultendpunct}{\mcitedefaultseppunct}\relax
\EndOfBibitem
\bibitem[Arbabi \latin{et~al.}(2015)Arbabi, Horie, Bagheri, and
  Faraon]{arbabi2015dielectric}
Arbabi,~A.; Horie,~Y.; Bagheri,~M.; Faraon,~A. \emph{Nature nanotechnology}
  \textbf{2015}, \emph{10}, 937--943\relax
\mciteBstWouldAddEndPuncttrue
\mciteSetBstMidEndSepPunct{\mcitedefaultmidpunct}
{\mcitedefaultendpunct}{\mcitedefaultseppunct}\relax
\EndOfBibitem
\bibitem[Chen \latin{et~al.}(2018)Chen, Zhu, Sanjeev, Khorasaninejad, Shi, Lee,
  and Capasso]{chen2018broadband}
Chen,~W.~T.; Zhu,~A.~Y.; Sanjeev,~V.; Khorasaninejad,~M.; Shi,~Z.; Lee,~E.;
  Capasso,~F. \emph{Nature nanotechnology} \textbf{2018}, \emph{13},
  220--226\relax
\mciteBstWouldAddEndPuncttrue
\mciteSetBstMidEndSepPunct{\mcitedefaultmidpunct}
{\mcitedefaultendpunct}{\mcitedefaultseppunct}\relax
\EndOfBibitem
\end{mcitethebibliography}

\begin{suppinfo}
\beginsupplement

\begin{figure}[b]
\centering 
\includegraphics[width=1\textwidth]{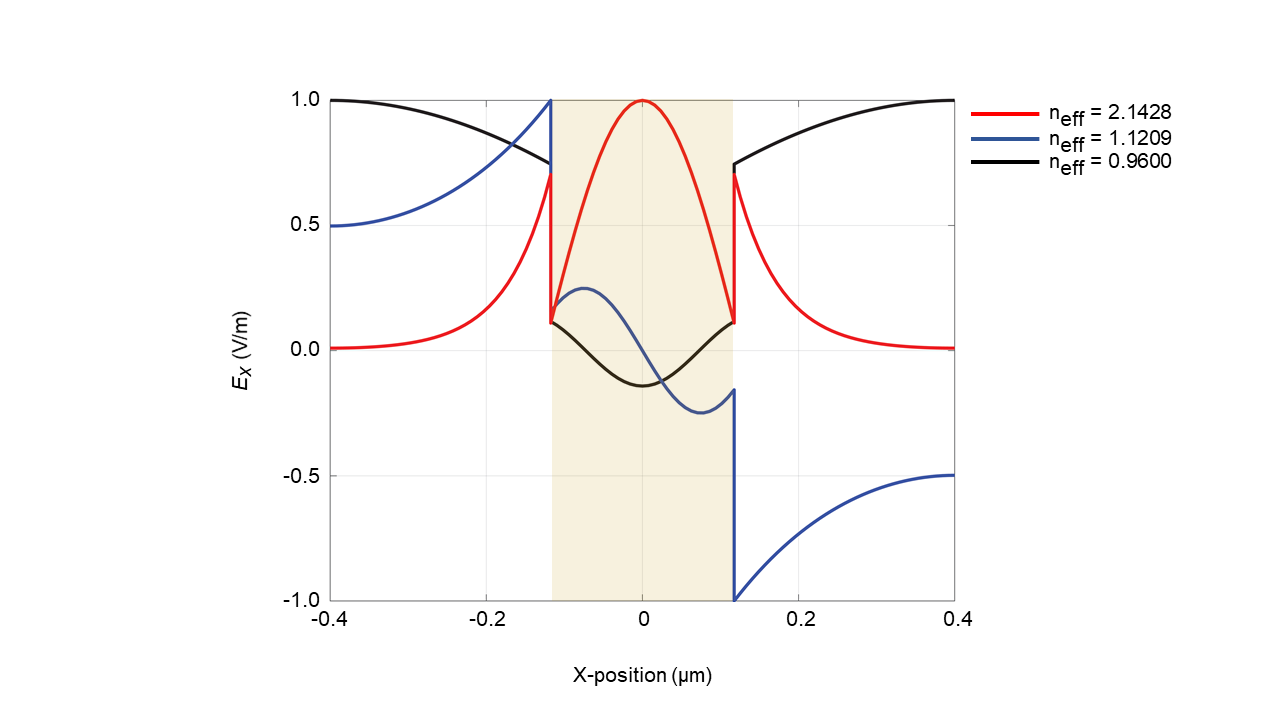}
\caption{Eigenmodes for the unit cell design proposed in Figure~\ref{fig:DWSG_symmetric_phy}. The beat length corresponds to the interference between the lowest order mode (red curve) with $n_\text{eff, A}$=2.14 and the radiation mode (black curve) with $n_\text{eff, B}$=0.96. } \label{fig:modes_symmetric}
\end{figure}

\begin{figure}[b]
\centering 
\includegraphics[width=1\textwidth]{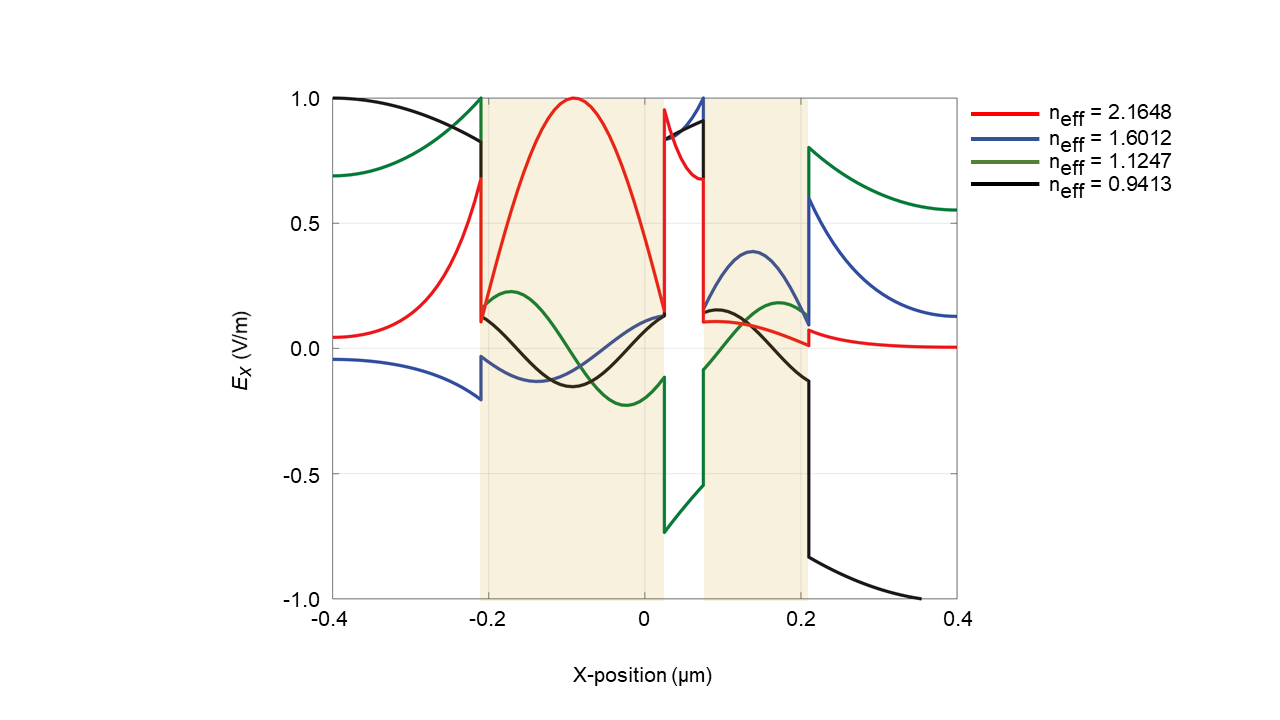}
\caption{Eigenmodes for the unit cell design proposed in Figure~\ref{fig:DWSG_asymmetric_des}. The beat length corresponds to the interference between the lowest order mode (red curve) with $n_\text{eff, A}$=2.16 and the higher order even guided mode (green curve) with $n_\text{eff, B}$=1.12. } \label{fig:modes_slot}
\end{figure}

\begin{figure}[b]
\centering 
\includegraphics[width=1\textwidth]{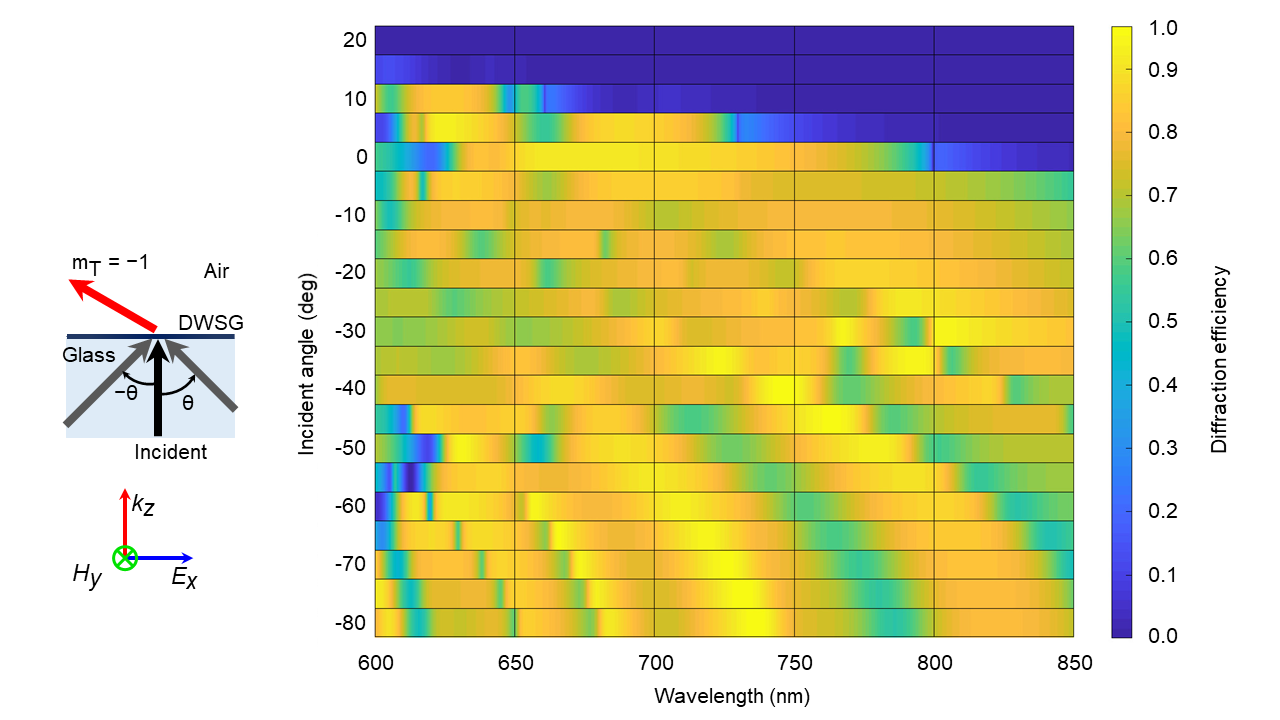}
\caption{Diffraction efficiency plot for the asymmetric one-dimensional slot-waveguide array design presented in Figure~\ref{fig:DWSG_asymmetric_des} for an incident beam from a variable angle of $\pm\theta$ with respect to the $z$-axis.} \label{fig:anglescan}
\end{figure}

\begin{figure}[b]
\centering 
\includegraphics[width=1\textwidth]{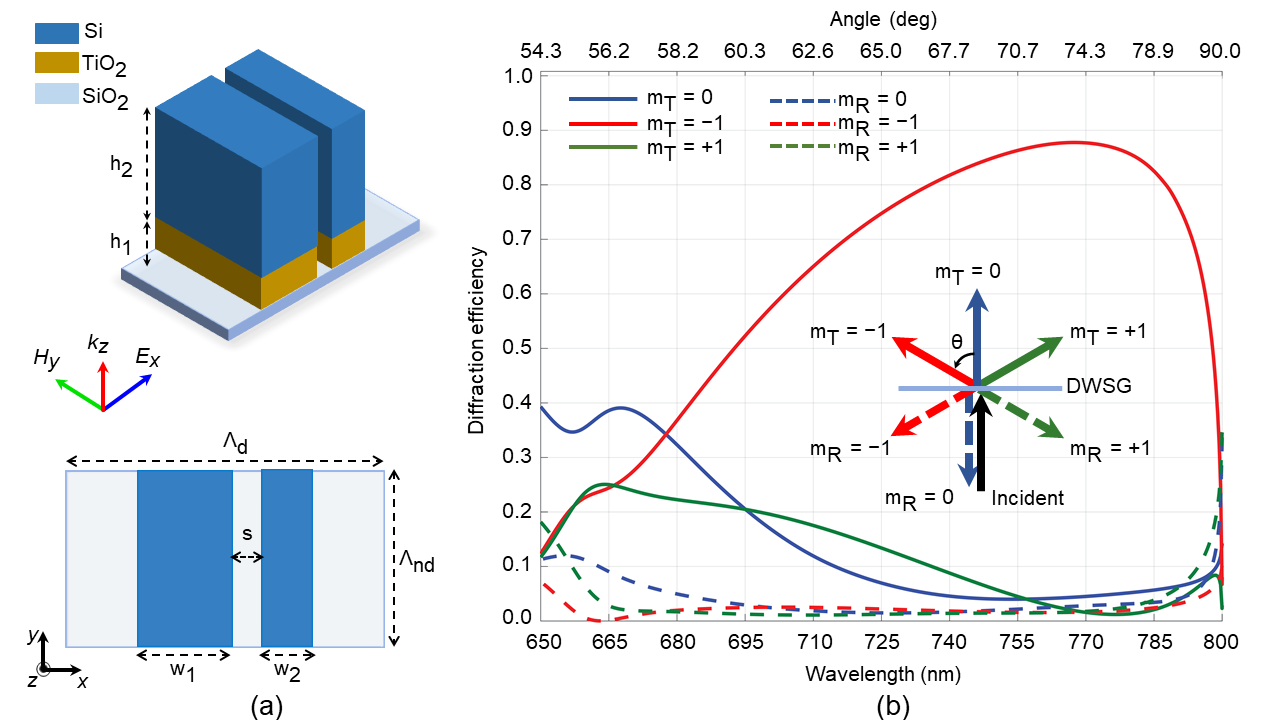}
\caption{Design of a unit cell comprising a one-dimensional, asymmetric and bi-layer slot-waveguide and its grating efficiency. (a) Perspective and top view. (b) Diffraction efficiency as a function of wavelength. The design parameter values are, in nm : $\Lambda_\text{d}=800,~\Lambda_\text{nd}=350,~w_\text{1}=165,~w_\text{2}=120,~h_\text{1}=135,~h_\text{2}=250,~\text{and}~ s=50$. A 3 dB bandwidth of 105 nm, ranging from 693 nm to 798nm, is obtained with a maximum efficiency of 87.5\% at 768 nm.} 
\label{fig:DWSG_asymmetric_bilayer}
\end{figure}

\begin{figure}[b]
\centering 
\includegraphics[width=1\textwidth]{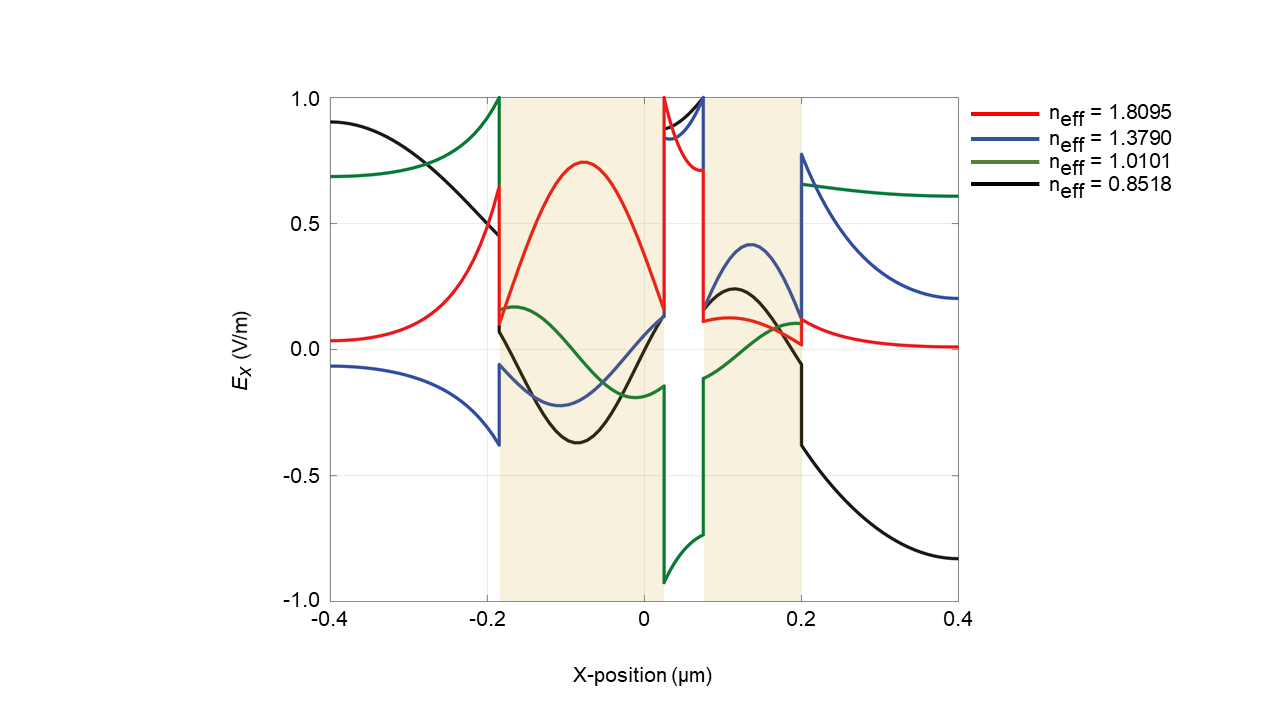}
\caption{Eigenmodes for the unit cell design proposed in Figure~\ref{fig:DWSG_asymmetric_des_2D}. The beat length corresponds to the interference between the lowest order mode (red curve) with $n_\text{eff, A}$=1.80 and the higher order even guided mode (green curve) with $n_\text{eff, B}$=1.01. } \label{fig:modes_ellipse}
\end{figure}

\begin{figure}[b]
\centering 
\includegraphics[width=1\textwidth]{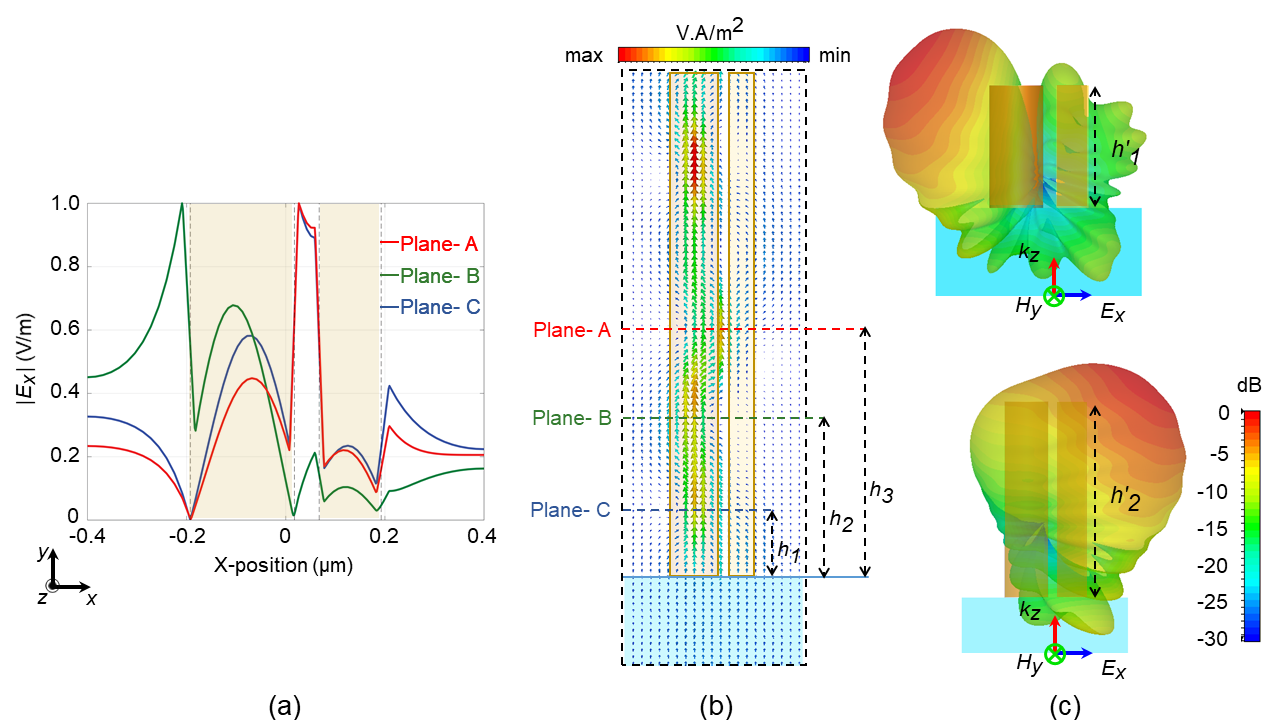}
\caption{Physical mechanism behind the scattering element presented in Figure~\ref{fig:DWSG_asymmetric_des_2D}. The half-beat length value is found to be $\sim$ 440 nm, for $\lambda=640$ nm, both from our 2-D mode solver ($n_\text{eff, A}=1.80, n_\text{eff, B}=1.01$) and 3-D full-wave simulation results. The horizontal cutting-plane A, B, and C are defined at a height of $h_{1}=310~\text{nm}, h_{2}=750~\text{nm},~\text{and}~h_{3}=1190~\text{nm}, \text{respectively}.$ (a) $| E_{x}|$ component on the intersection lines between Plane-A, B, and C with Plane-D. (b) Absolute Poynting vector plot on Plane-D. (c) Directional radiation pattern for an open-ended waveguide section with height $h'_{1}=475$ nm (top) and $h'_{2}=950$ nm (bottom).
} \label{fig:DWSG_asymmetric_phy_2D}
\end{figure}

\begin{figure}[b]
\centering 
\includegraphics[width=1\textwidth]{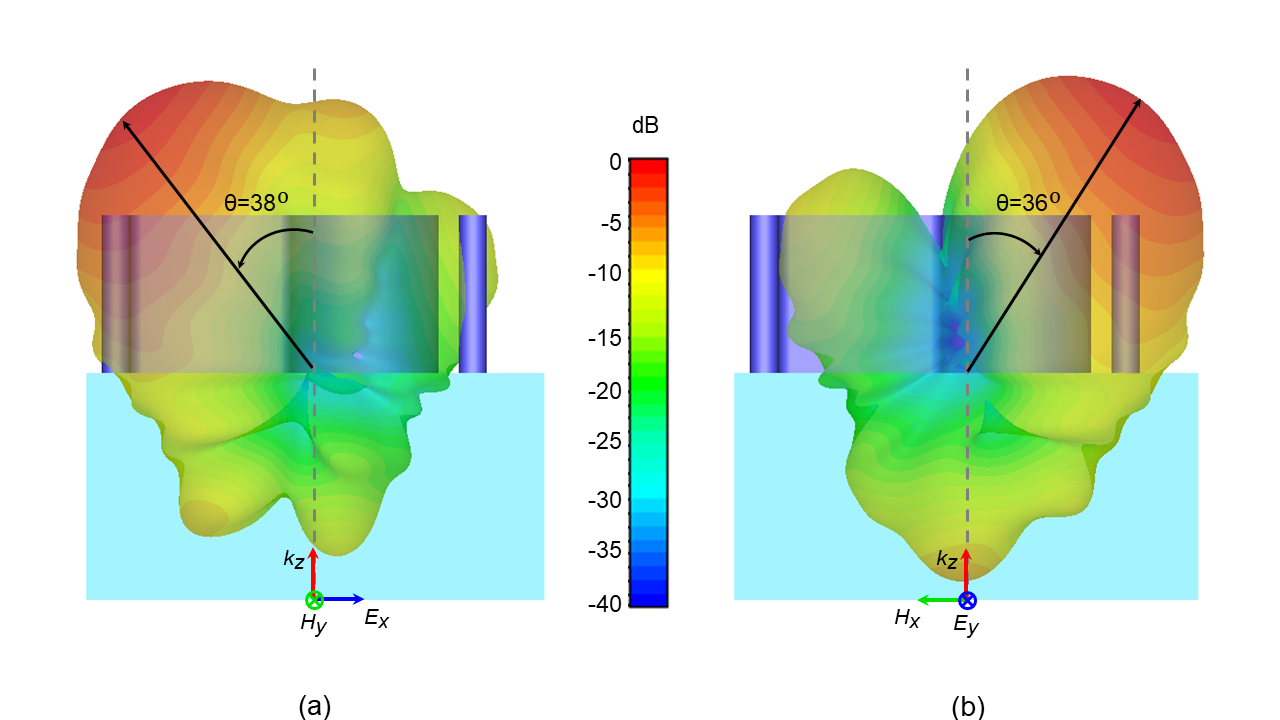}
\caption{Radiation pattern of the PBS-DWSG unit cell design proposed in Figure~\ref{fig:DWSG_PBS_des} for (a) $x$-polarized incident light, and (b) $y$-polarized incident light. The black-arrows points in the direction of maximum scattering for each polarization incident.} \label{fig:DWSG_PBS_phy}
\end{figure}

\end{suppinfo}

\end{document}